\documentclass[twocolumn,preprintnumbers,amsmath,amssymb,prl]{revtex4-1}
\usepackage[pdftex]{graphicx}
\usepackage{dcolumn}% Align table columns on decimal point
\usepackage{bm}% bold math
\usepackage{color}
\usepackage{rotating}
\usepackage{ulem} 
\usepackage{lineno}
\begin{document}

\title{
Magnetothermal transport in ultraclean single crystals of Kitaev magnet $\alpha$-RuCl$_3$
}

\author{Y. Xing$^1$}
\author{R. Namba$^2$}
\author{K. Imamura$^2$}
\author{K. Ishihara$^2$}
\author{S. Suetsugu$^3$}
\author{T. Asaba$^3$}
\author{K. Hashimoto$^2$}
\author{T. Shibauchi$^2$}
\author{Y. Matsuda$^3$}
\author{Y. Kasahara$^4$}
\email{kasahara.yuichi.437@m.kyushu-u.ac.jp}

\affiliation{$^1$College of New Energy and Materials, China University of Petroleum, Beijing 102249, People's Republic of China}
\affiliation{$^2$Department of Advanced Materials Science, University of Tokyo, Chiba 277-8561, Japan}
\affiliation{$^3$Department of Physics, Kyoto University, Kyoto 606-8502, Japan}
\affiliation{$^4$Department of Physics, Kyushu University, Fukuoka 819-0395, Japan}

\date{\today}% 

%\pacs{}  

%% Abstract

\begin{abstract}
	
The layered honeycomb magnet $\alpha$-RuCl$_3$ has emerged as a promising candidate for realizing a Kitaev quantum spin liquid. Previous studies have reported oscillation-like anomalies in the longitudinal thermal conductivity and half-integer quantized thermal Hall conductivity above the antiferromagnetic critical field $H_c$, generating significant interest. However, the origins of these phenomena remain contentious due to strong sample dependence. Here we re-examine the magnetothermal transport properties using recently available ultra-pure $\alpha$-RuCl$_3$ single crystals to further elucidate potential signatures of the spin liquid state. Our findings reveal that while anomalies in thermal conductivity above $H_c$ persist even in ultraclean crystals, their magnitude is significantly attenuated, contrary to the quantum oscillations hypothesis. This suggests that the anomalies are likely attributable to localized stacking faults inadvertently introduced during magnetothermal transport measurements. The thermal Hall conductivity exhibits a half-quantized plateau, albeit with a narrower width than previously reported. 
This width reduction can be understood through two distinct mechanisms: sample-dependent magnetic critical fields that influence the lower boundary of the plateau region, and the decoupling between chiral Majorana edge currents and phononic thermal transport that determines the upper boundary. 
These results indicate that structural imperfections exert a substantial influence on both the oscillation-like anomalies and quantization effects observed in magnetothermal transport measurements of $\alpha$-RuCl$_3$.

\end{abstract}
\maketitle

%% Introduction
\section*{Introduction}

Quantum spin liquid (QSL) is an exotic state of matter in which quantum fluctuations prevent conventional long-range magnetic order, and hosts many fascinating phenomena, such as  fractionalized excitations and topological orders \cite{Balents10}. The Kitaev model on a two-dimensional (2D) honeycomb lattice is a novel example of an exactly solvable QSL state that exhibits itinerant Majorana fermions and localized $Z_2$ fluxes (visons) \cite{Kitaev06}. In magnetic fields, these emergent quasiparticles give rise to non-Abelian anyons, which can be useful for topological quantum computing.   Among candidate materials, the layered honeycomb magnet $\alpha$-RuCl$_3$ \cite{Plumb14} has been intensively studied as one of the most promising candidates \cite{Takagi19,Trebst2022}.   Although a zigzag antiferromagnetic (AFM) order appears below the N\'{e}el temperature $T_{\rm N}\sim7$\,K in zero field, the signatures of spin fractionalization have been reported by various experiments, suggesting that $\alpha$-RuCl$_3$ is in proximity to a Kitaev QSL \cite{Do17,Widmann19,Sandilands15,Banerjee16,Banerjee18,Balz19}. More importantly, the AFM order is suppressed by in-plane magnetic fields, and the field-induced quantum disordered (FIQD) state above $\mu_0H_c\sim8$\,T is proposed to be a Kitaev QSL state.  

In the FIQD state of $\alpha$-RuCl$_3$, two remarkable observations have been reported by thermal transport experiments.   One is the half-integer quantized thermal Hall (HIQTH) effect, which has been confirmed by several groups \cite{Kasahara18,Yamashita20,Bruin21,Yokoi21}. The 2D thermal Hall conductance per honeycomb plane $\kappa_{xy}^{\rm 2D}$ shows a quantized plateau behavior as a function of $H$ and is close to half the value of the thermal conductance quantum $K_0=\frac{\pi^2k_B^2}{3h}T$.   Moreover, a finite thermal Hall conductivity is observed even when the magnetic field is applied parallel to the in-plane thermal current, a phenomenon known as the planar thermal Hall effect. This effect exhibits half-quantization \cite{Yokoi21}.  Notably, the field-angular variation of the HIQTH conductance has the same sign structure as the topological Chern number expected for the Kitaev QSL \cite{Yokoi21}.   These results provide direct evidence for chiral Majorana modes at the sample edge and non-Abelian anyons in the bulk \cite{Kitaev06}.   The half-quantized planar thermal Hall effect in $\alpha$-RuCl$_3$ exhibits intriguing temperature-dependent behavior. The quantization is observed in a limited temperature range of 3.5-6.5\,K.  At lower temperatures, $\kappa_{xy}^{\rm 2D}$ is reduced to values smaller than the quantized thermal conductance. 
%
%conductance is 
%
%reduced to be smaller than $K_0$.  
This phenomenon has been interpreted in the context of  decoupling between chiral Majorana edge modes and phonons \cite{Vinkler18,Ye18}.  However, the universality of these observations remains contested.   Some research groups report no evidence of the plateau or quantization \cite{Lefrancois21}, instead proposing bosonic magnon origins for the thermal Hall effect \cite{Czajka23}.   On the other hand, recent experimental observations have revealed two significant concurrent phenomena when a magnetic field is applied parallel to the $b$-axis: a sign reversal in the planar thermal Hall effect and the closing of an energy gap, resulting in the formation of a Dirac cone \cite{Tanaka22,Imamura23}. This simultaneous manifestation of these distinct and remarkable physical processes provides compelling evidence supporting the hypothesis that Majorana fermions are responsible for the observed thermal Hall effect \cite{Tanaka22,Imamura23}.  The other remarkable observation is the oscillation-like anomalies of the longitudinal thermal conductivity $\kappa$ \cite{Czajka21,Suetsugu22,Bruin22,Lefrancois23,Zhang23_PRM,Zhang24,Zhang23_arxiv}. This anomalies have been reproduced, but interpretations of their origin remain highly controversial. Both intrinsic and extrinsic origins have been discussed, such as charge-neutral fermions in a putative QSL state \cite{Czajka21}, which may be associated with the Kitaev QSL state \cite{Zhang23_arxiv}, and a sequence of magnetic phase transitions induced by stacking faults \cite{Suetsugu22,Bruin22,Lefrancois23}.  

Sample dependence likely accounts for the conflicting results and interpretations of the thermal Hall effect and thermal conductivity measurements.  In fact, $\alpha$-RuCl$_3$ is known to be highly susceptible to stacking faults that can be introduced during the crystal growth process and sample handling \cite{Cao16}.   The half-integer quantized plateau has only been confirmed for samples grown by the Bridgman method, but there is no such report for the samples grown by the chemical vapor transport (CVT) method. Moreover, even for samples grown by the Bridgman method, the HIQTH effect is observed only in samples with higher longitudinal thermal conductivity than a certain threshold value \cite{Kasahara22}. As for the thermal conductivity oscillations, some groups reported the enhancement of the oscillation amplitude in samples with lower longitudinal thermal conductivity \cite{Bruin22,Lefrancois23,Namba24}, while the others reported the opposite trend \cite{Zhang23_PRM}.   Thus, the situation calls for further thermal transport studies on samples with an extreme cleanliness. 

In this paper, we revisit the magnetothermal transport properties of $\alpha$-RuCl$_3$ using ultraclean samples that have been recently synthesized \cite{Namba24}.   We found that anomalous behavior manifesting as oscillatory-like features is observed in the magnetic field dependence of the thermal conductivity, $\kappa(H)$, above $H_c$, but their magnitude is significantly suppressed in the ultraclean samples.   This contradicts with the hypothesis of quantum oscillations, indicating that the oscillatory features arise from a sequence of magnetic transitions induced by unavoidable stacking faults.   The thermal Hall conductivity exhibits a half-quantized plateau, but its width is strongly reduced in our samples with higher $T_{\rm N}$ and $H_c$. We discuss these results in terms of 
sample-dependent magnetic critical fields and mean free path. 
%

%% Experiment
 
%Single crystals of $\alpha$-RuCl$_3$ were grown using a two-step crystal growth method combining with the CVT and sublimation processes \cite{Namba24}.   
%
%We used crystals with no anomaly associated with the magnetic transition at 10-14\,K due to the stacking faults. We carefully selected as-grown crystals that were thin plate-like and long enough in the direction of heat current $\mathbf{q}$. 
%
%The in-plane longitudinal thermal conductivity $\kappa$ and the thermal Hall conductivity $\kappa_{xy}$ were measured by the standard steady-state method. For the measurements of $\kappa_{xy}$, magnetic fields ${\bm H}$ were applied along the $a$-axis. The details of the structural and physical properties of our samples are described in Ref.\,\cite{Namba24}. 

Previous measurements of thermal Hall conductivity primarily used a configuration with $\mathbf{q}\parallel a$, determining $\kappa_{xy}$ under the assumption that $\kappa(\mathbf{q}\parallel a)=\kappa(\mathbf{q}\parallel b)$ \cite{Kasahara18,Yamashita20,Bruin21,Yokoi21}. However, this equivalence is not guaranteed by crystal symmetry \cite{Kurumaji23}. Therefore, measurements with $\mathbf{q}\parallel b$ are crucial for conclusively validating the half-integer thermal Hall quantization in $\alpha$-RuCl$_3$. In this study, we measured the thermal Hall conductivity with the thermal current applied along the $b$-axis ($\mathbf{q}\parallel b$).

% Fig.1
%\begin{figure}[h]
\begin{figure}[t]
 	\begin{center}
		\includegraphics[width=1.0\linewidth]{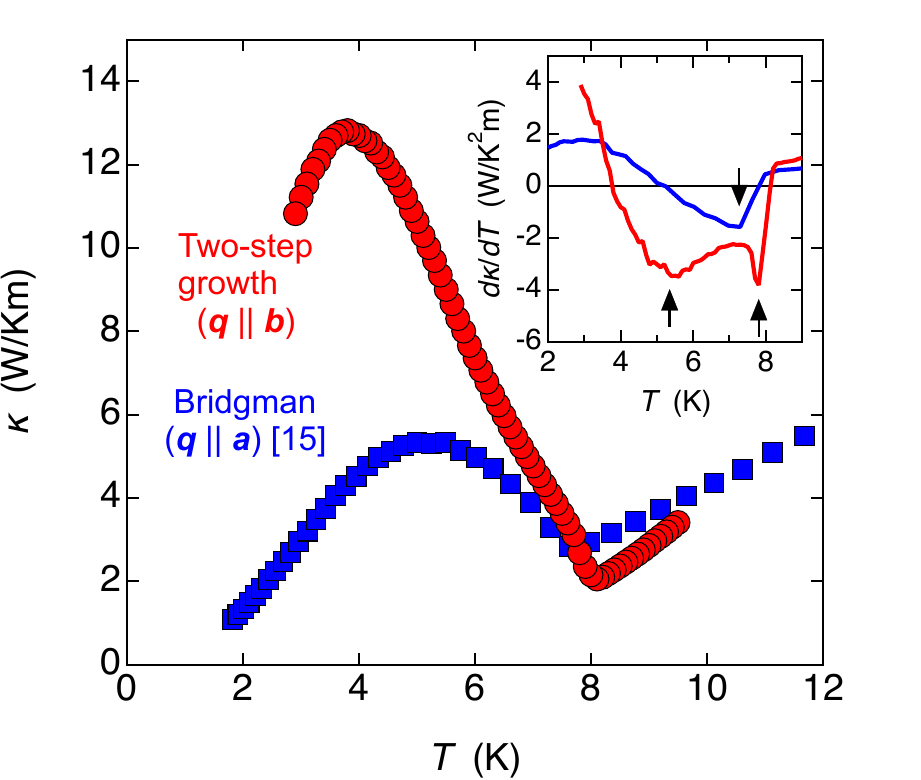}
		\caption{
		{\bf Temperature dependence of the longitudinal thermal conductivity.} 
		Temperature dependence of the longitudinal thermal conductivity $\kappa$ for the ultraclean (two-step sublimation) sample (red) is compared with those of the Bridgman sample (blue) \cite{Yokoi21}. The heat current $\mathbf{q}$ is applied along the $b$-axis ($\mathbf{q}\parallel b$) for the sublimation sample and along the $a$-axis ($\mathbf{q}\parallel a$) for the Bridgman sample. 
		The inset shows the temperature dependence of $d\kappa/dT$.   The arrows indicates minima in $d\kappa/dT$. 
		}
 	\end{center}
 \end{figure}

%Fig. 2
\begin{figure}[t]
%\begin{figure}[h]
	\begin{center}
		\includegraphics[width=1.0\linewidth]{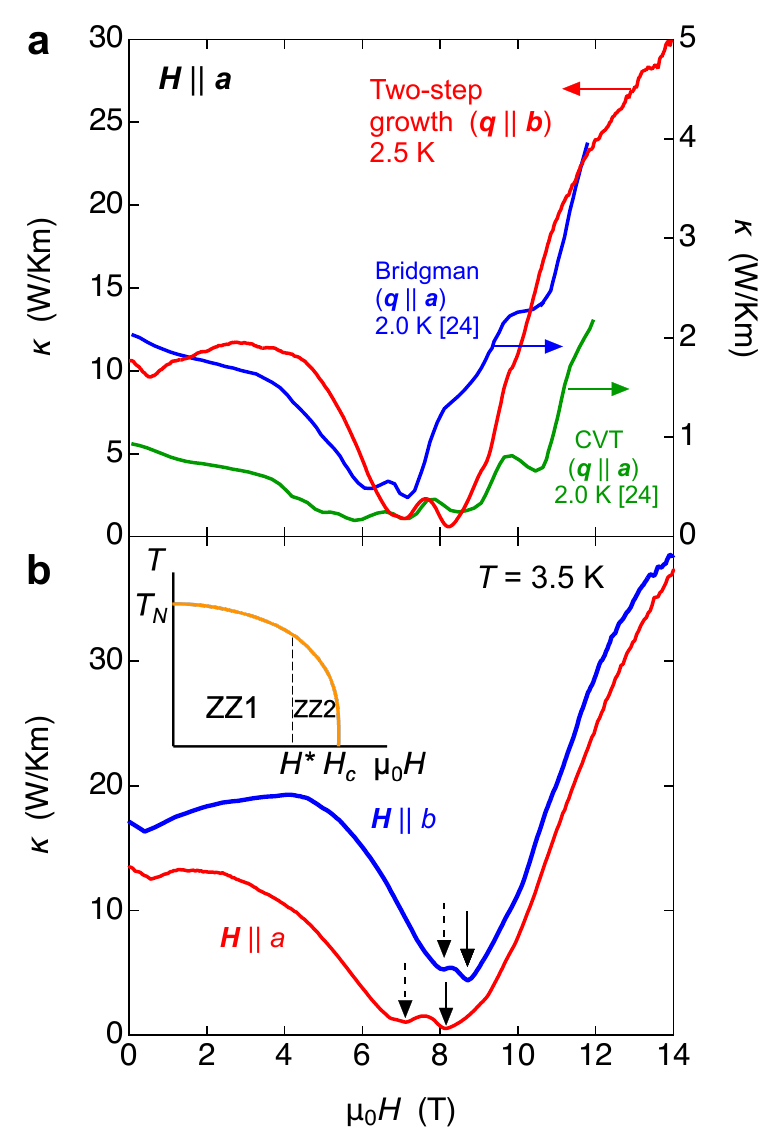}
		\caption{
		{\bf Field dependence of $\kappa$.} 
		{\bf a} Field dependence of $\kappa$ with applied magnetic field along $a$-axis for ultraclean (two-step sublimation) sample (red, left axis) is compared with those of the Bridgman sample (blue, right axis) and CVT sample (green, right axis) \cite{Bruin22}. 
Note that the scale of the left axis is six times larger than that of the right axis. 
		{\bf b} The $H$-dependence of $\kappa$ at 3.5\,K for $\bm{H}\parallel a$ and $\bm{H}\parallel b$.  Sharp minima are caused by the magnetic transitions.   The data for $\bm{H}\parallel b$ are vertically shifted for clarity.   The solid (dashed) arrows represent the phase boundary between 
the low-field zigzag AFM phase (zz1) and intermediate zigzag AFM phase with a modified zigzag pattern along the layers (zz2) 
at $H_c$($H^\ast$). The inset shows a schematic phase diagram in in-plane magnetic fields.  
			}
	\end{center}
\end{figure}

%% Results and Discussions
\section*{Results and Discussion}
\subsection*{Temperature dependence of $\kappa$}

Figure\,1 shows the temperature dependence of $\kappa$ in zero field. Data of Bridgman sample are also plotted for comparison \cite{Yokoi21}.    The thermal conductivity exhibits a sharp kink anomaly at around $T_{\rm N}=7.8$\,K, which is higher than that of the Bridgman sample ($T_{\rm N}=7.4$\,K), as previously reported \cite{Namba24}. Here, $T_{\rm N}$ was determined from the minima in $d\kappa/dT$ (Fig.\,1, inset). 
The variation of $T_{\rm N}$ between samples appears to depend on the degree of disorder, such as point defects \cite{Zhang23_PRM,Namba24,Imamura24}, as the magnetic interaction is very sensitive to the local crystal symmetry in $\alpha$-RuCl$_3$ \cite{Kim24}. This is in contrast to stacking faults that induce an additional magnetic transition above 10\,K while leaving $T_{\rm N}$ largely unperturbed \cite{Cao16,Namba24}. 
The peak of $\kappa(T)$ below $T_{\rm N}$ is more than twice as large as that of the Bridgman sample and is the highest among those reported so far. This demonstrates the superior quality of the samples with minimal disorder and stacking faults \cite{Namba24}. Hereafter, the two-step sublimation sample will be referred to as the ultraclean sample. Although $\kappa(T)$ of the ultraclean sample looks similar to that of the Bridgman sample, a clear difference can be seen by taking the temperature derivative of $\kappa(T)$, $d\kappa(T)/dT$.   Below $T_{\rm N}$, two dip structures appear in $d\kappa(T)/dT$ at about 5.5 and 7.8\,K, indicating two-step enhancement of $\kappa$ below $T_N$, in contrast to a single dip at about 7.4\,K for the Bridgman sample (arrows in the inset of Fig.\,1).   In $\alpha$-RuCl$_3$, heat is carried by both magnetic excitations and phonons, $\kappa=\kappa_{\rm mag}+\kappa_{\rm ph}$.   It has been suggested that $\kappa_{\rm ph}$ is the dominant contribution, and the peak of $\kappa$ appears as a result of an enhanced phonon mean free path due to the suppression of spin-phonon scattering upon entering the AFM phase \cite{Hentrich18,Leahy17}.  
The observation of two-step enhancement in $\kappa(T)$ below $T_{\rm N}$, in the absence of any magnetic or structural phase transitions, suggests either a non-negligible  contribution from $\kappa_{\rm{mag}}$ or the presence of unconventional magnetic excitation spectra that uniquely modulate the phonon mean free path in the zigzag-ordered state of $\alpha$-RuCl$_3$. This behavior deviates from conventional magnetic systems and warrants detailed consideration of the interplay between magnetic and lattice degrees of freedom. 
This result, which has never been reported before, suggests the importance of high-quality samples for investigating the intrinsic magnetothermal transport properties of $\alpha$-RuCl$_3$.

% Fig. 3
\begin{figure}[t]
%\begin{figure}[h]
	\begin{center}
		\includegraphics[width=1.0\linewidth]{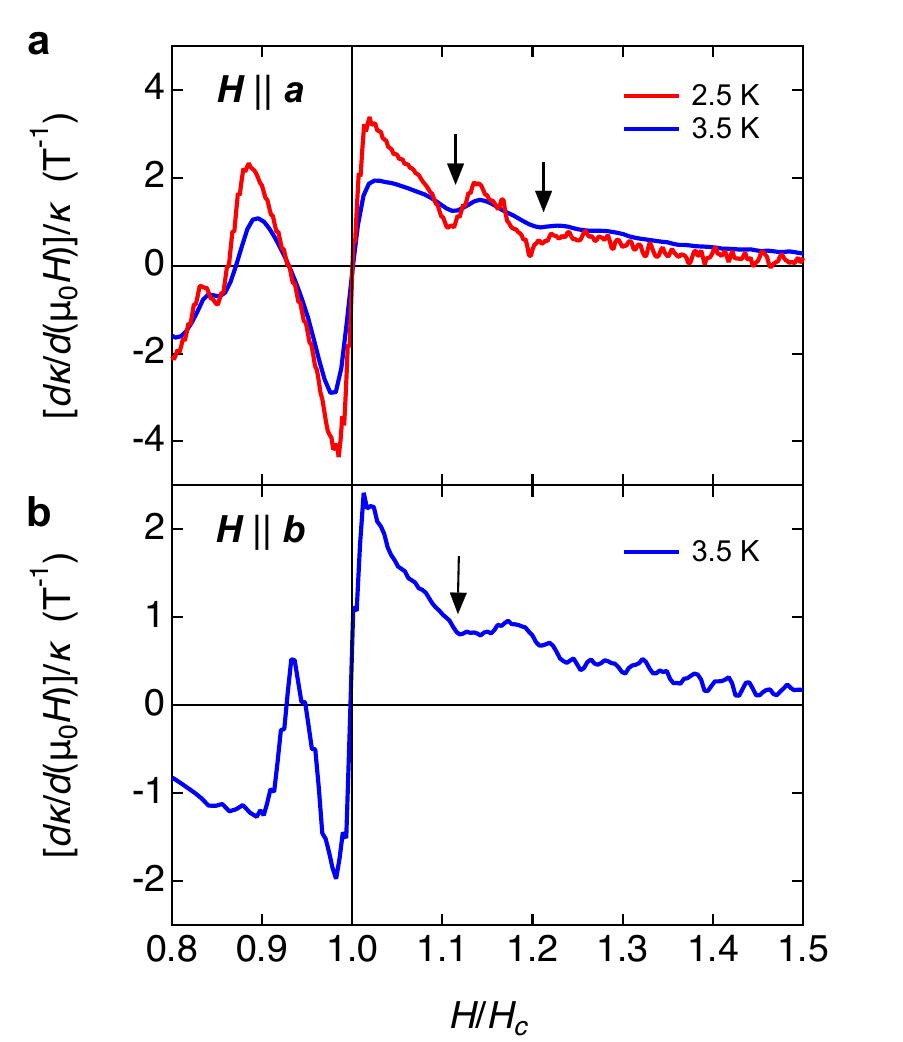}
		\caption{
		{\bf Anomalies of field derivative of $\kappa$ above critical fields.}  
		The normalized field derivative of $\kappa$, $[d\kappa/d(\mu_0H)]/\kappa$, plotted as a function of the normalized field $H/H_c$ for field applied along ({\bf a}) $a$- and ({\bf b}) $b$-axes ($H_c=8.2$\, T and 8.7\,T for $a$- and $b$-axes, respectively). The arrows represent the locations of the minima in $[d\kappa/d(\mu_0H)]/\kappa$. The arrows indicate the well-defined minima in the FIQD state, which likely originate from the oscillation-like features of $\kappa(H)$ reported in the Bridgman and CVT samples. 
		}
	\end{center}
\end{figure}

\subsection*{Oscillation-like anomalies in $\kappa(H)$}

Figure\,2a depicts the field dependence of $\kappa$ for $\bm{H}\parallel a$ at 2.5\,K in the ultraclean sample.  For comparison, the data from samples grown using  the Bridgman and CVT methods are also shown \cite{Bruin22}.   At high magnetic fields above about 8\,T, $\kappa(H)$ increases rapidly, and its magnitude becomes much larger than those of Bridgman and CVT crystals.  Remarkably, in the ultraclean sample, despite the significantly higher thermal conductivity in the FIQD state,  the oscillatory features are much less prominent compared to the Bridgman and CVT samples.  We will discuss this later.  

We here discuss the distinctions in the zigzag AFM state between our ultraclean crystal and previously studied Bridgman-grown crystals.  Figure\,2b compares $\kappa(H)$ measured for two orthogonal in-plane field orientations, $\bm{H}\parallel a$ and $\bm{H}\parallel b$.  As depicted in Fig.\,2b, $\kappa(H)$ exhibits two distinct minima for both field orientations, indicated by dashed and solid arrows. The solid arrow corresponds to $H_c$, marking the transition from the AFM phase to the FIQD state. The dashed arrow denotes the transition field $H^\ast$  from the low-field zigzag AFM phase to 
another zigzag AFM phase with a modified zigzag pattern along the layers 
(Fig.\,2b, inset).  It is noteworthy that the anomaly at $H^\ast$  has not been clearly resolved for $\bm{H}\parallel b$ in Bridgman-grown samples. This is primarily because $H^\ast$  in such samples is very close to $H_c$ ($H_c-H^\ast\lesssim0.2$\,T), making the distinction between these two transitions difficult. 
Higher $T_N$ results in higher $H_c$, $H^\ast$, and the clear separation of $H^\ast$ and $H_c$ in our ultraclean crystals, allowing for a more precise characterization of the magnetic phase diagram and the intermediate zigzag AFM phase, which may have significant implications for understanding the magnetic properties of $\alpha$-RuCl$_3$.

The oscillation-like anomalies are significantly reduced but still discernible even in our ultraclean crystal.  This can be seen clearly by plotting the field derivative of $\kappa(H)$, $[d\kappa/d(\mu_0H)]/\kappa$, as a function of the normalized field $H/H_c$ in Fig.\,3a, b. In the FIQD state above $H_c$, minima are observed for both $\bm{H}\parallel a$ and $\bm{H}\parallel b$, as indicated by the arrows.  The amplitude of these minima increases with decreasing temperature for $\bm{H}\parallel a$, consistent with the previously reported oscillation-like anomalies.  This indicates that the oscillation-like anomalies above $H_c$ persist even in the ultraclean crystals.   However, the fact that significant suppression of the magnitude of anomalies in the ultraclean crystals with higher thermal conductivity obviously contradicts the hypothesis of quantum oscillations.   Instead, these anomalies are likely attributed to a series magnetic transitions induced by stacking faults.   
Although no evidence of stacking faults was detected in the ultraclean samples prior to the experiments, we cannot rule out the possibility that stacking faults were introduced during the experiments.  In the present thermal conductivity setup, uniaxial stain on crystal is inevitably present \cite{Yokoi21}, which may induce stacking faults.   These stacking faults could potentially lead to a series of secondary magnetic transition, resulting in oscillation-like anomalies in $\kappa(H)$.  The anomalies in $\kappa(H)$ in the FIQD state occur at remarkably similar normalized field values ($H/H_c$) for both $\bm{H} \parallel a$ and $\bm{H} \parallel b$, as indicated by the arrows in Fig.\,3a, b.  This close correspondence in the positions of the anomalies across different field orientations suggests that these anomalies are intrinsically linked to the magnetic anisotropy of $\alpha$-RuCl$_3$.   It should be noted that even a series of secondary magnetic phases have been reported to exhibit magnetic anisotropy similar to that of the zigzag AFM phase \cite{Bruin22}. While further precise measurements on ultraclean crystals are necessary to definitively determine the origin of the oscillatory-like anomalies, the present results suggest an extrinsic origin for these features. The observed suppression of anomaly magnitude in higher-conductivity samples, coupled with the consistent occurrence at similar normalized field values across different field orientations, points towards structural imperfections such as local stacking faults as a probable cause.

%Fig.4
\begin{figure*}[t]
%\begin{figure}[h]
	\begin{center}
		%\includegraphics[width=0.7\linewidth]{Fig4.pdf}
		%\hspace*{-0.1\linewidth}
		\includegraphics[width=0.9\linewidth]{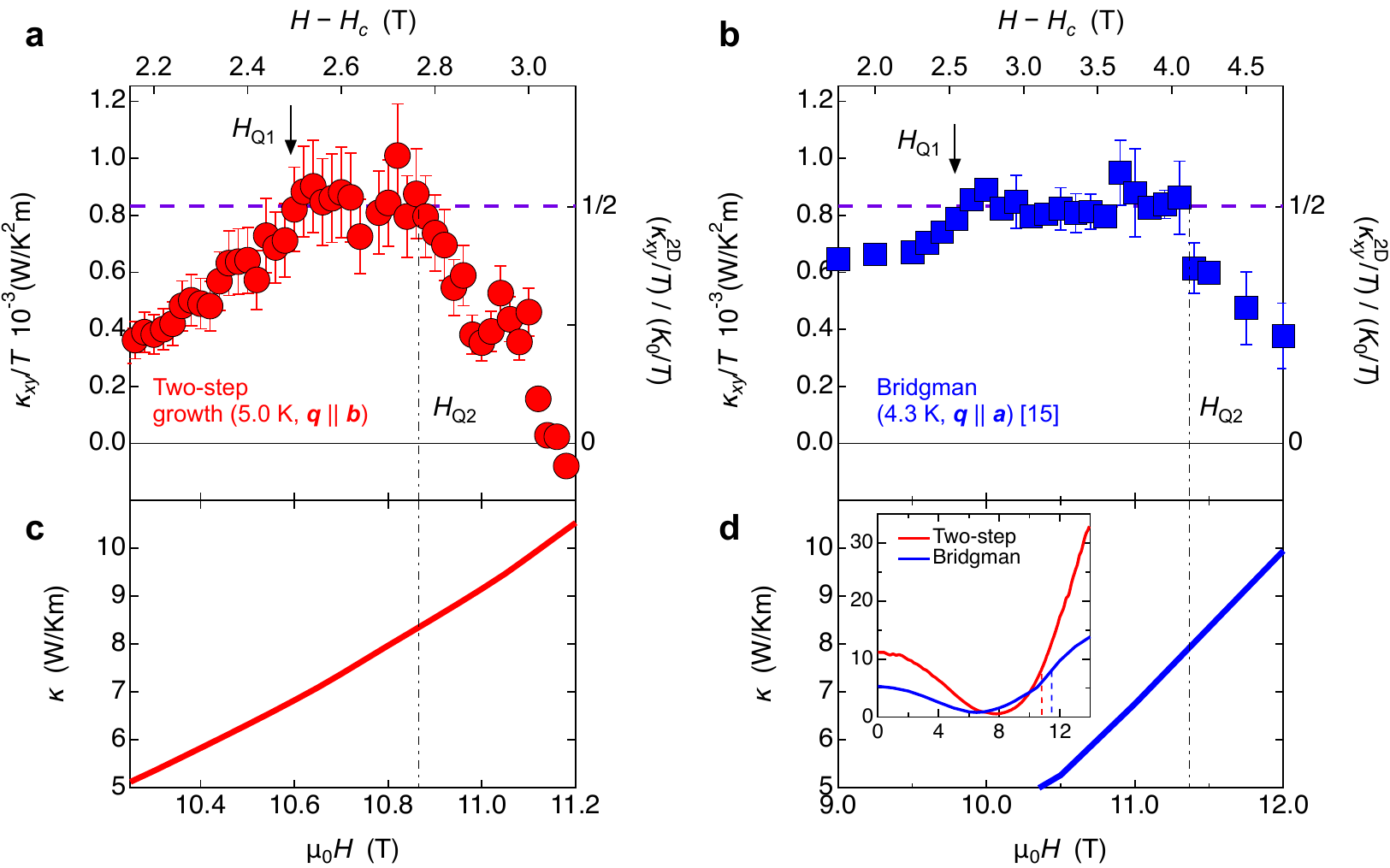}
		\caption{
		{\bf Half-integer thermal Hall conductance plateau.}  
		Thermal Hall conductivity divided by $T$, $\kappa_{xy}/T$, and thermal conductivity $\kappa$ of ({\bf a}, {\bf c}) the ultraclean sample and ({\bf b}, {\bf d}) the Bridgman sample. {\bf a}, {\bf b} Field dependence of $\kappa_{xy}/T$ in the in-plane fields parallel to the $a$-axis for ({\bf a}) the ultraclean sample at 5.0\,K and ({\bf b}) the Bridgman sample at 4.3\,K \cite{Yokoi21}. In the right axes, thermal Hall conductance is shown in units of the quantum thermal conductance $K_0=(\pi^2/3)(k_\mathrm{B}^2/h)T$. The dashed violet line indicates the half-integer quantization. The arrows and vertical dash-dotted lines indicate the lower and upper boundaries of the plateau, $H_{Q1}$ and $H_{Q2}$, respectively. 
		{\bf c}, {\bf d} Field dependence of $\kappa$ for ({\bf c}) the ultraclean sample at 5.0\,K and ({\bf d}) the Bridgman sample at 4.3\,K \cite{Yokoi21}. The inset shows the same plot as ({\bf c}, {\bf d}) over a wide field range. The red (blue) dashed line represents $H_{Q2}$ of the ultraclean (Bridgman) sample. 
		}
	\end{center}
\end{figure*}

\subsection*{Half-integer quantization of the planer thermal Hall effect}

Next, we present the results of the thermal Hall effect on the ultraclean sample.   Figure\,4a depicts the thermal Hall conductivity $\kappa_{xy}$ with $\bm{H}\parallel-a$ at 5.0\,K.   The dashed violet line represents the HIQTH conductance, $\kappa_{xy}^{\rm 2D}/T=\frac{1}{2}(K_0/T)$, where $\kappa_{xy}^{\rm 2D}=\kappa_{xy}d$ with the interlayer distance $d$. Similar to the previous reports, $\kappa_{xy}$ of the ultraclean sample exhibits a plateau behavior as a function of $H$, and its value is very close to the HIQTH conductance \cite{Yokoi21,Kasahara18,Yamashita20,Bruin21}.   We note that this represents the first observation of the HIQTH effect with $\mathbf{q}\parallel b$, whereas all previous reports \cite{Yokoi21,Kasahara18,Yamashita20,Bruin21} utilized $\mathbf{q}\parallel a$.  

Our present results demonstrate that $\kappa(\mathbf{q}\parallel a)$ is indeed very close to $\kappa(\mathbf{q}\parallel b)$. This result is supported by the data shown in Fig.\,1, where $\kappa(\mathbf{q}\parallel b)$ for the ultraclean sample is comparable to $\kappa(\mathbf{q}\parallel a)$ for the Bridgman sample above $T_{\rm N}$. 
In fact, the near sample-independence of $\kappa$'s magnitude above $T_{\rm{N}}$ is observed, indicating that the mean free path of heat carriers is limited by the magnetic fluctuations rather than disorders above $T_{\rm{N}}$ \cite{Hirobe17}. These results suggest that a comparison of $\kappa(\mathbf{q}\parallel a)$ and $\kappa(\mathbf{q}\parallel b)$ provides insight into the intrinsic thermal conductivity anisotropy in $\alpha$-RuCl$_3$.  
Thus, while measurements with $\mathbf{q}\parallel a$ and $\mathbf{q}\parallel b$ using the same sample would be ideal for a direct comparison, our results nonetheless provide additional support for the existence of the HIQTH conductance plateau in $\alpha$-RuCl$_3$. 

The observed field range for quantization $(H_{Q1}<H<H_{Q2})$ exhibits differences between ultraclean and Bridgman samples (Fig.\,4a, b). Quantitatively, the plateau field range in the Bridgman sample is approximately 2.5 times greater than that of the ultraclean sample that possesses higher thermal conductivity.  Furthermore, $H_{Q2}$ shifts to lower fields, while $H_{Q1}$ increases in the ultraclean sample, suggesting different mechanisms governing $H_{Q1}$ and $H_{Q2}$.  It has been theoretically proposed that critical factor in observing quantized thermal Hall plateaus in Kitaev QSLs is the coupling between chiral Majorana edge modes and phonons \cite{Vinkler18,Ye18}. Notably, deviations from the quantized thermal Hall value occur at low temperatures, where the phonon mean free path $\ell_{\rm{ph}}$ approaches the crystal size.  In this ballistic phonon regime, decoupling between phonons and edge currents becomes significant.  In $\alpha$-RuCl$_3$, this decoupling phenomenon becomes more pronounced even under the application of a magnetic field. Specifically, in the range of magnetic fields highlighted in (Fig.\,4c, d) (see also the inset of Fig.\,4d), a dramatic increase in thermal conductivity corresponding to the increase in phonon mean free path with increasing magnetic field strength was observed.  Consequently, disorder-induced reduction of the phonon mean free path enhances phonon-edge coupling, leading to an expected increase of $H_{Q2}$ for samples with higher disorder.  The similar crystal sizes and coinciding threshold $\kappa\approx8$\,W/Km suggest that phonon-edge current decoupling initiates when the phonon mean free path approaches crystal dimensions, aligning with theoretical predictions. This threshold $\kappa$ value for quantization supports theoretical work highlighting the crucial role of interactions between chiral Majorana edge current and phonons. The lower critical field $H_{Q1}$ appears well above $H_c$, with substantial differences observed between Bridgman and ultraclean crystals. Notably, despite these differences, we find that the field interval between $H_{Q1}$ and $H_c$ ($H_{Q1}-H_c$) remains similar ($\sim2.5$\,T) for both Bridgman and ultraclean crystals, as shown in Fig.\,4a, b. While the underlying mechanism requires further investigation, this similarity suggests that $H_{Q1}$ is closely related to the underlying magnetic interactions that determine $H_c$. Although further systematic studies using crystals with controlled disorder are necessary to confirm these conclusions, our results provide valuable insights into the quantization phenomenon in $\alpha$-RuCl$_3$.

%% Summary
%In summary, we conducted measurements of the longitudinal thermal conductivity and thermal Hall conductivity in ultraclean $\alpha$-RuCl$_3$ samples, synthesized using a two-step sublimation method. Our findings reveal discrepancies with previously reported results from magnetothermal transport measurements on  $\alpha$-RuCl$_3$ grown by Bridgman and CVT methods.
%We observed oscillation-like anomalies in the longitudinal thermal conductivity of our ultraclean samples. However, the magnitude of these anomalies was significantly suppressed compared to crystals grown by other methods. This observation suggests that these anomalies likely stem from extrinsic effects, possibly due to secondary magnetic transitions induced by local stacking faults unintentionally present even in ultraclean samples. This also underscores the importance of careful sample preparation and characterization in the study of this compound. Furthermore, we detected a half-integer quantized thermal Hall conductance plateau in the ultraclean crystal. Interestingly, the width of this half-integer quantized plateau in the ultraclean sample is reduced compared to that grown by the Bridgman method. 
Our comparative analysis %of these systems 
suggests that the phonon-edge coupling and sample-dependent magnetic interactions play key roles in determining the upper and lower field bounds of the half-quantized plateau, respectively. 
%
%This study emphasizes 
These results emphasize 
the significance of sample quality and growth methods in understanding the intrinsic properties of $\alpha$-RuCl$_3$ and their implications for Kitaev QSL physics.

%% Methods
%\section*{Methods}
%\subsection*{Sample growth}
\vskip\baselineskip
\noindent
{\bf Methods}\\
{\bf Sample growth}\\
Single crystals of $\alpha$-RuCl$_3$ were grown using a two-step crystal growth method combining with the chemical vapor transport and sublimation process \cite{Namba24}. We used crystals with no anomaly associated with the magnetic transition at 10-14\,K due to the stacking faults. We carefully selected as-grown crystals that were thin plate-like and long enough in the direction of heat current $\mathbf{q}$. The details of the structural and physical properties of our samples are described in Ref.\,\cite{Namba24}.

%\subsection*{Thermal transport measurements}
\noindent
{\bf Thermal transport measurements}\\
The in-plane longitudinal thermal conductivity $\kappa$ and the thermal Hall conductivity $\kappa_{xy}$ were measured by the standard steady-state method. For the measurements of $\kappa_{xy}$, magnetic fields $\bm{H}$ were applied along the $a$-axis. 

%% Data Availability
%\section*{Data Availability}
\vskip\baselineskip
\noindent
{\bf Data availability}\\
%All data needed to evaluate the conclusions are present in the paper. 
The data that support the findings of this study are available from the corresponding author upon reasonable request.

%% Acknowledgments
%
%\begin{acknowledgments}
%\section*{Acknowledgements}
\vskip\baselineskip
\noindent
{\bf Acknowledgements}\\
We thank E.-G. Moon, T. Kurumaji, and J. Nasu for insightful discussions. This work is supported by Grants-in-Aid for Scientific Research (KAKENHI) (Nos.\,JP23K13060, JP23H00089, JP24H00965, and JP24K21529) and Grants-in-Aid for Scientific Research on Innovative Areas ``Quantum Liquid Crystals" (No.\,JP19H05824) from Japan Society for the Promotion of Science (JSPS), and CREST (JPMJCR19T5) and PRESTO (JPMJPR2354) from Japan Science and Technology Agency (JST). 
%\end{acknowledgments}

%% Author Contributions
%\section*{Author Contributions}
\vskip\baselineskip
\noindent
{\bf Author contributions}\\
T.S., Y.M., and Y.K. conceived and supervised the study. 
Y.X., S.S., T.A., and Y.K. performed the thermal transport measurements. 
R.N., K.Imamura, K.Ishihara, K.H., and T.S. synthesized the single crystals.  
Y.X. and Y.K. analyzed the data with inputs from S.S., T.A., T.S., and Y.M. 
All authors discussed the results and contributed to writing the manuscript.

%% Competing Interests 
%\section*{Competing Interests}
\vskip\baselineskip
\noindent
{\bf Competing interests}\\
The authors declare no competing interests. 

%% Figure

%% Reference

\end{document}